\documentclass[
reprint,
 amsmath,amssymb,
 aip,
]{revtex4-2}

\usepackage{graphicx}% Include figure files
\usepackage{dcolumn}% Align table columns on decimal point
\usepackage{bm}% bold math
\usepackage{enumitem}
\usepackage{float}
\usepackage{mathtools}  % allows \matrix
\usepackage{esvect} % allows \vec
\usepackage{empheq}  % allows \empheq
\usepackage{xcolor}

\DeclareMathOperator{\sinc}{sinc}

\newcommand{\affSFU}{Department of Physics, Simon Fraser University, Burnaby, British Columbia, Canada V5A 1S6}
\newcommand{\affUWece}{Department of Electrical and Computer Engineering, University of Waterloo, Ontario, Canada N2L 3G1}
\newcommand{\affUW}{Department of Physics and Astronomy, University of Waterloo, Ontario, Canada N2L 3G1}
\newcommand{\affMcM}{Department of Engineering Physics, McMaster University, Ontario, Canada L8S 4L7}
\usepackage{caption}
\usepackage[most]{tcolorbox}  % allows \framebox
\usepackage[makeroom]{cancel}  % allows \cancel
\usepackage{nccmath}  % allows \fleqn
\usepackage{braket}
\usepackage{tikz} % arrowed description
\usepackage{tabularx}				%tables that wrap around the text nicely
\usepackage{array}					%tables that center and wrap around text
\usepackage{hhline} %  cline is incompatible with coloring tables, so use hhline instead
\usepackage{physics}
\usepackage{siunitx}
\usepackage{multirow} % allows \multirow in tables
\usepackage{hyperref}% add hypertext capabilities
\begin{document}
\raggedbottom
\preprint{APS/123-QED}

\title{Hybrid Entanglement Based Long Distance Quantum Communication\\using Thin Film Lithium Niobate Photonic Chip}% Force line breaks with \\

\author{Sungeun Oh}
\email{SungeunOh68@gmail.com}
\affiliation{\affSFU}

\author{Arsalan Khan}
\affiliation{\affUWece}

\author{Hamish Johnson}
\affiliation{\affSFU}

\author{Eva Pearson}
\affiliation{\affUW}

\author{Marko Pocrnjic}
\affiliation{\affSFU}

\author{Leila Mehvar}
\affiliation{\affMcM}

\author{Saeed Oghbaey}
\affiliation{\affMcM}

\author{Chang-qing Xu}
\affiliation{\affMcM}

\author{Sasan V. Grayli}
\affiliation{\affSFU}
\affiliation{\affUWece}

\author{Thomas Jennewein}
\email{thomas\_jennewein@sfu.ca}
\affiliation{\affSFU}
\affiliation{\affUW}

\date{\today}

\begin{abstract}

Large-scale quantum networks will increasingly rely on photonic integrated circuits (PICs), which offer compact, mass-producible platforms for scalable quantum networking. At the same time, such networks are expected to adopt hybrid architectures that combine low-loss optical fiber links with long-distance free-space channels. Here, we propose and numerically investigate a thin-film lithium niobate (TFLN)-based PIC design for a compact hybrid entangled-photon source suited to such networks. The source generates non-degenerate photon pairs, with a $\SI{790}{\nano\meter}$ photon encoded in polarization for free-space transmission and a $\SI{1550}{\nano\meter}$ photon encoded in time bins for optical fiber transmission. We deliberately restrict the design to simple device components rather than maximizing efficiency, and examine how far such a circuit can support long-distance quantum communication. Our simulations indicate that, despite these losses, the high photon-pair generation rate of  $\sim\SI{100}{\mega\hertz}$ at only $\SI{35}{\micro\watt}$ of guided pump power, enabled by TFLN makes it a suitable platform for entangled photon sources in large-scale quantum networks. This work also provides a conservative baseline and design guidelines for the future fabrication of integrated hybrid entangled photon sources, as a step toward terrestrial and satellite-based quantum communication.
\end{abstract}

\maketitle

%%%%%%%%%%%%%%%%%%%%%%%%%%

\section{Introduction}
Quantum information technologies are expected to significantly enhance our computing and communication capabilities while enabling entirely new functions that classical systems cannot achieve. The key to these technologies lies in quantum entanglement. For instance, quantum entanglement between photon pairs shared between distant systems can support distributed quantum computing, teleportation-based communication, and enhanced measurement precision across sensor networks \cite{Kimble2008}. Quantum entanglement is also utilized in various Quantum Key Distribution (QKD) protocols, enabling users to share quantum information securely over long distances. Combined with rapidly growing concerns over cybersecurity, governments worldwide are making substantial investments in quantum communication research, recognizing quantum-secure networking as a strategic asset for future global networking structure. Over the past two decades, quantum communication technologies based on both optical fiber and free-space links have advanced rapidly through extensive studies and field demonstrations \cite{Pelet2023, Yin2017}. However, it is increasingly recognized that a hybrid approach combining both types of links will play a crucial role in establishing a large-scale quantum network due to several practical considerations.
In addition to the hybrid approach, photonic integration offers a route to reducing the footprint, alignment requirements, and number of interfaces between separately implemented generation and encoding components, which is also important for large-scale quantum networks. A big advantage of PICs lies in their ability to be mass-produced, which will be an economical solution for expanding quantum networking infrastructure. However, integration alone does not guarantee efficient photon-pair generation. Widely used photonic platforms, such as silicon ($Si$) and silicon nitride ($Si_3N_4$), possess centrosymmetric crystal structures and therefore typically rely on third-order nonlinear processes, such as four-wave mixing (FWM), for photon-pair generation. Depending on the platform and device design, achieving sufficiently high photon-pair generation rates using FWM can be challenging, particularly in applications requiring long-distance transmission.
The importance of high photon-pair generation rates is illustrated by satellite-based quantum communication missions such as QEYSSat, for which link-budget simulations have suggested link losses of approximately 43 dB \cite{Bourgoin_2013}. While still small compared to those in optical fiber over the same distance, such substantial losses motivate the development of efficient photon-pair sources to ensure sufficient detection rates at the receiving end.
In contrast, lithium niobate ($\mathrm{LiNbO}_3$) exhibits strong second-order optical nonlinearity, enabling efficient photon-pair generation via spontaneous parametric down-conversion (SPDC). This material system provides an attractive route toward achieving the high photon-pair generation rates required for long-distance quantum communication and the establishment of large-scale quantum networks. Therefore, we propose a thin-film lithium niobate (TFLN) photonic chip-based design for a hybrid entangled photon source (EPS) for long-distance quantum communication. 

\newpage
%%%%%%%%%%%%%%%%%%%%%%%%%%

\section{Hybrid Channels for Large-Scale Quantum Networks}

\begin{figure}[htbp!]
	\centering
	\includegraphics[width=\columnwidth]{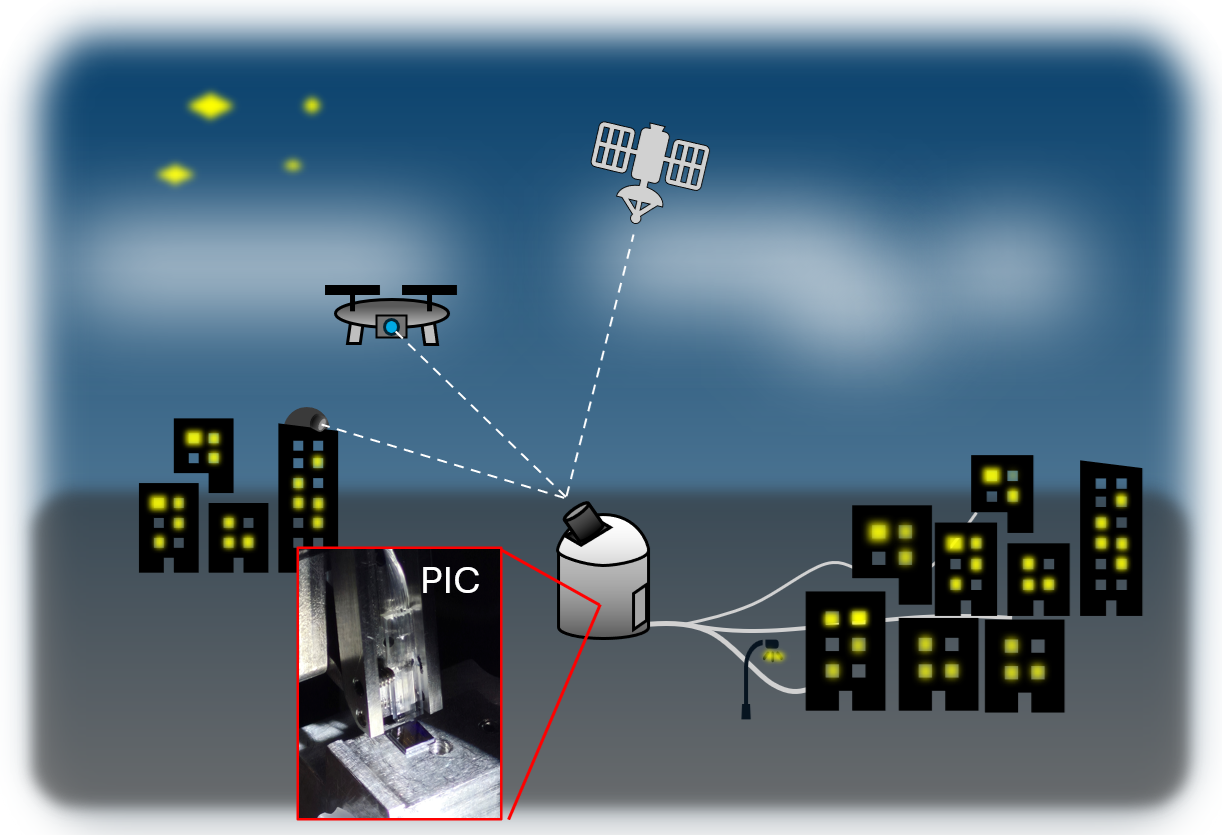}
	\footnotesize
	\caption{Photonic Integrated Circuit (PIC) as an entangled photon source.
}\label{HybridNetwork}
\end{figure}

While optical fiber links offer the advantage of utilizing already established telecommunications infrastructure, they attenuate optical signals exponentially with distance, limiting the rate of direct photon transmission.
Even the most developed and optimized telecom-band fibers currently achieve losses of approximately 0.3 dB/km, making the use of repeaters inevitable for long-distance transmission. Some early-stage demonstrations of large-scale quantum networks have been conducted, connecting multiple cities over thousands of kilometers through the use of multiple classical repeater nodes\cite{Chen2021}. Since the classical information handled by these nodes may be exposed to external parties, extending such architectures to metropolitan-scale or larger networks will require substantial effort and infrastructure investment. Meanwhile, quantum-memory-based networking has been proposed and extensively studied as a viable route for establishing long-distance quantum links \cite{Knaut2024, Yu2020, Liu2024}. However, not only does this approach face complex technical challenges, such as interfacing between photons and qubits, but also it has yet to surpass the long-range transmission achieved by free-space links.

Relying solely on free-space links is not always the ideal solution either, as they come with their own challenges. Unlike optical fiber-based links, free-space connectivity have in fact shown successful long-distance quantum communication, ranging from tens to thousands of kilometers. However, free-space transmission requires line-of-sight connection, and therefore, it remains sensitive to weather, and is subject to background light that can reduce the signal-to-noise ratio, followed by an increase in quantum bit error rate \cite{Yastremski2025}. 
While challenges in both fiber-based and free-space links are being addressed, combining the two offers a practical route toward realizing global quantum networks.

In principle, both polarization and time-bin encoding can be used over either free-space or optical fiber channels. In practice, however, each channel favors a particular encoding scheme, and a hybrid quantum network combining the two is most efficiently implemented by matching a well-suited encoding to each channel. Time-bin encoding is well suited to optical fiber transmission, as quantum information encoded in temporally separated modes is largely insensitive to polarization fluctuations induced by variations in fiber birefringence  \cite{Brendel1999}. In free-space links, conventional time-bin analyzers rely on phase-stable interferometers, in which turbulence-induced wavefront distortions can degrade interference visibility  \cite{jin_Demonstration_2018}. Polarization encoding, in contrast, is robust in free-space links and generally offers better preservation of the encoded information through atmospheric channels \cite{scalcon_CrossEncoded_2022}. 
A hybrid EPS can combine these complementary encoding schemes by entangling the polarization degree of freedom of one photon with the time-bin degree of freedom of the other. Implementing such a source on a photonic chip requires not only the integration of these encoding schemes but also the preservation of the coherence necessary for entanglement generation, as well as efficient coupling of the two photons to their respective channels.

%%%%%%%%%%%%%%%%%%%%%%%%%%%%

\section{Non-degenerate Photon Pairs on TFLN Platform}
Here, we propose and numerically investigate a TFLN-based photonic circuit for generating non-degenerate hybrid-entangled photon pairs, with the near-infrared photons encoded in polarization and the telecom-band photons encoded in time bins. The use of non-degenerate photon pairs is motivated by the distinct operational advantages of free-space and fiber-optic channels: photons at $\SI{1550}{\nano\meter}$ benefit from low attenuation in standard telecom fibers and compatibility with existing metropolitan infrastructure, while photons at  $\SI{790}{\nano\meter}$ are readily compatible with silicon single-photon detectors and established free-space quantum optics. Recent advances towards the proposed hybrid architecture include on-chip coherent conversion between photonic degrees of freedom \cite{Feng2016}, as well as development of TFLN-based sources of non-degenerate photon-pair \cite{Babel2025} and polarization-entangled photons \cite{Kim2026}. 

The design platform comprises a submicrometer lithium niobate device layer on a low index SiO$_2$ insulator, supported by a substrate; waveguides are patterned into the lithium niobate \cite{Wang2018,Zhao2020}. SPDC process allows converting one pump photon into a signal-idler pair through the second order susceptibility $\chi^{(2)}$. Energy conservation requires 
\begin{equation}
    \omega_p=\omega_s+\omega_i
    \label{eq:tfln-energy}
\end{equation}
where $p$, $s$, and $i$ denote the pump, signal, and idler, respectively. Here, a pulsed pump centred at approximately $\SI{523.5}{\nano\meter}$ is selected to generate pairs centred at $\SI{790}{\nano\meter}$ and $\SI{1550}{\nano\meter}$. In the absence of phase matching, the probability amplitudes associated with photon-pair generation at different positions along the waveguide can interfere destructively, reducing the overall generation rate. Periodic poling alternates the sign of the nonlinear coefficient along the waveguide and therefore, an appropriate period allows these contributions to add constructively \cite{Spillane2007,Fejer1992}. For the first-order quasi-phase matching (QPM), the residual mismatch is
\begin{equation}
    \Delta\beta = \beta_p-\beta_s-\beta_i-\dfrac{2\pi}{\Lambda},
    \label{eq:tfln-qpm}
\end{equation}
where $\Lambda$ is the poling period and $\beta$=$2\pi n/\lambda$ are the propagation constants of the guided modes. In on-chip waveguides, while Eq.~\eqref{eq:tfln-energy} remains unchanged, one must consider the modified effective indices ($n\rightarrow n_{\text{eff}}$) of the guided modes and the required poling period resulting from the waveguide geometry and the cladding substrate. Here, we consider a periodically poled rib waveguide region with a total film thickness of $\SI{150}{\nano\meter}$, an etch depth of $\SI{70}{\nano\meter}$ with a $\ang{60}$ sidewall angle and a top width of $\SI{1}{\micro\meter}$. From modal analysis using Finite Difference Eigenmode (FDE) simulations (Ansys, Lumerical), the obtained effective indices are $n_{eff,p} = 1.977$, $n_{eff,s} = 1.782$ and $n_{eff,i} = 1.543$. Setting $\Delta\beta=0$ in Eq.~\eqref{eq:tfln-qpm}, a first-order poling period of $\Lambda=1.9\mu m$ is required for quasi-phase matched SPDC. 
The group indices are also obtained from the FDE simulations, yielding $n_{g,p}=2.511$, $n_{g,s}=2.217$ and $n_{g,i}=1.806$. For a periodically poled region of length $L=\SI{3}{\milli\meter}$, full spectral bandwidths of the photon pairs at half maximum (FWHM) are calculated as $\Delta\nu\approx\SI{216} {\giga\hertz}$ by Taylor-expanding to the first order.
\begin{equation}
    \Delta\nu\approx\dfrac{0.886c}{L\left|n_{g,s}-n_{g,i}\right|}
    \label{eq:bandwidth}
\end{equation}

For ideal first-order periodic poling with a $50\%$ duty cycle, the effective nonlinear coefficient is $d_{\text{eff}}=2d_{33}/\pi\approx17.2pm/V$, where $d_{33}=27$pm/V is the largest nonlinear tensor element of LiNbO$_3$, corresponding to a type-0 interaction with pump, signal, and idler in the TE-polarized modes. Leveraging tight optical confinement and strong second-order nonlinearity, TFLN waveguides achieve exceptional photon-pair generation rates via SPDC. The efficiency of this process is primarily governed by the spatial overlap of the pump, signal, and idler modes \cite{Wang2018,Zhao2020}. In a scalar approximation for the selected polarizations, the effective mode overlap area is
\begin{equation}
    A_{\text{eff}}=\frac{\displaystyle\prod_{j=p,s,i}\int_{\text{all}}|E_j|^2\,dA}{\displaystyle\left|\int_{WG}E_pE_s^*E_i^*\,dA\right|^2},
    \label{eq:tfln-overlap}
\end{equation}
where $E_j$ denotes the transverse electric-field profile of each mode. The numerator accounts for the field normalization over the entire cross-section, while the denominator includes only the spatial overlap of the three modes within the waveguide. A smaller effective modal overlap area enhances the nonlinear interaction by increasing the spatial overlap of the interacting optical fields. From numerical integration of the FDE electric-field distributions, we obtained an effective interaction area of $A_{\text{eff}}=\SI{1.02}{\micro\meter}^2$.

Pulsed pump with a pulse width of $\tau_p=\SI{50}{\pico\second}$, and a repetition rate of $f_{\text{rep}}=\SI{1}{\giga\hertz}$ are assumed for the pump. Although the pump is pulsed, the quasi-continuous-wave (quasi-CW) approximation can still be used to estimate the pair-generation rate, provided that the pump spectral bandwidth is much narrower than the phase-matching acceptance bandwidth or equivalently, in the time domain, when the pump pulse duration $\tau_p$ is much longer than the temporal walk-off $\tau_{\mathrm{wo}}$ accumulated over the poling length $L$ due to group-velocity mismatch.
\begin{equation}
  \tau_p \gg \tau_{\mathrm{wo}} = \frac{L}{c}\,
  \max\left(\left|n_{g,p}-n_{g,s}\right|,\,\left|n_{g,p}-n_{g,i}\right|\right)
  \label{eq:walkoff}
\end{equation}
Using the group indices obtained from FDE simulations, we calculate a temporal walk-off of approximately $\tau_{\mathrm{wo}}=\SI{7.0}{\pico\second}$. The chosen pump pulse duration exceeds the walk-off by almost an order of magnitude, so the pump can be regarded as quasi-monochromatic over the phase-matching bandwidth, validating the quasi-CW treatment of the pair-generation rate.
For a lossless waveguide in the low-gain regime, the pair generation rate per unit signal frequency is given by \cite{Spillane2007}
\begin{equation}
    \frac{dR_{\text{pair}}}{d\nu_s}
    =\dfrac{2d_{\text{eff}}^2\omega_s\omega_iP_{\text{p}} L^2}
    {\epsilon_0 n_{\text{eff},p}n_{\text{eff},s}n_{\text{eff},i}
    A_{\text{eff}}}
    \sinc^2 \left(\frac{\Delta\beta L}{2}\right),
    \label{eq:tfln-spectral-rate}
\end{equation}
where $P_p$ is the average guided pump power and $\epsilon_0$ is the vacuum permittivity. 
The generated pair rate is obtained by integrating Eq.~\eqref{eq:tfln-spectral-rate} over the emission spectrum.
\begin{equation}
\begin{aligned}
    R_{\text{pair}}&=\int
    \frac{dR_{\text{pair}}}{d\nu_s}\,d\nu_s\\
    &\approx\dfrac{2d_{\text{eff}}^2\omega_s\omega_iP_{\text{p}} L}
    {\epsilon_0c^2n_{\text{eff},p}n_{\text{eff},s}n_{\text{eff},i}
    A_{\text{eff}}\left|n_{g,s}-n_{g,i}\right|}
    \label{eq:rate}
\end{aligned}
\end{equation}
Link-budget analyses for satellite-based quantum communication indicate that a photon-pair generation rate of $\sim\SI{100}{\mega\hertz}$ is required to overcome $\sim\SI{40}{\deci\bel}$ loss and background photon noise \cite{Bourgoin_2013,OhPaper1}. Solving Eq.~\eqref{eq:rate} for our $\SI{3}{\milli\meter}$ TFLN waveguide shows that this rate can be achieved with an average guided pump power of only $P_p\approx\SI{35}{\micro\watt}$.

%%%%%%%%%%%%%%%%%%%%%%%%%%

\section{Photonic Integrated Circuit for Hybrid Entanglement}\label{Section4}
Based on the waveguide modes and phase-matching conditions established in the previous section, we present a photonic integrated circuit (PIC) on X-cut TFLN that implements a compact hybrid-EPS for terrestrial fiber and satellite-based free-space quantum links.
\begin{figure*}[htbp!]
	\centering
    \includegraphics[width=\textwidth]{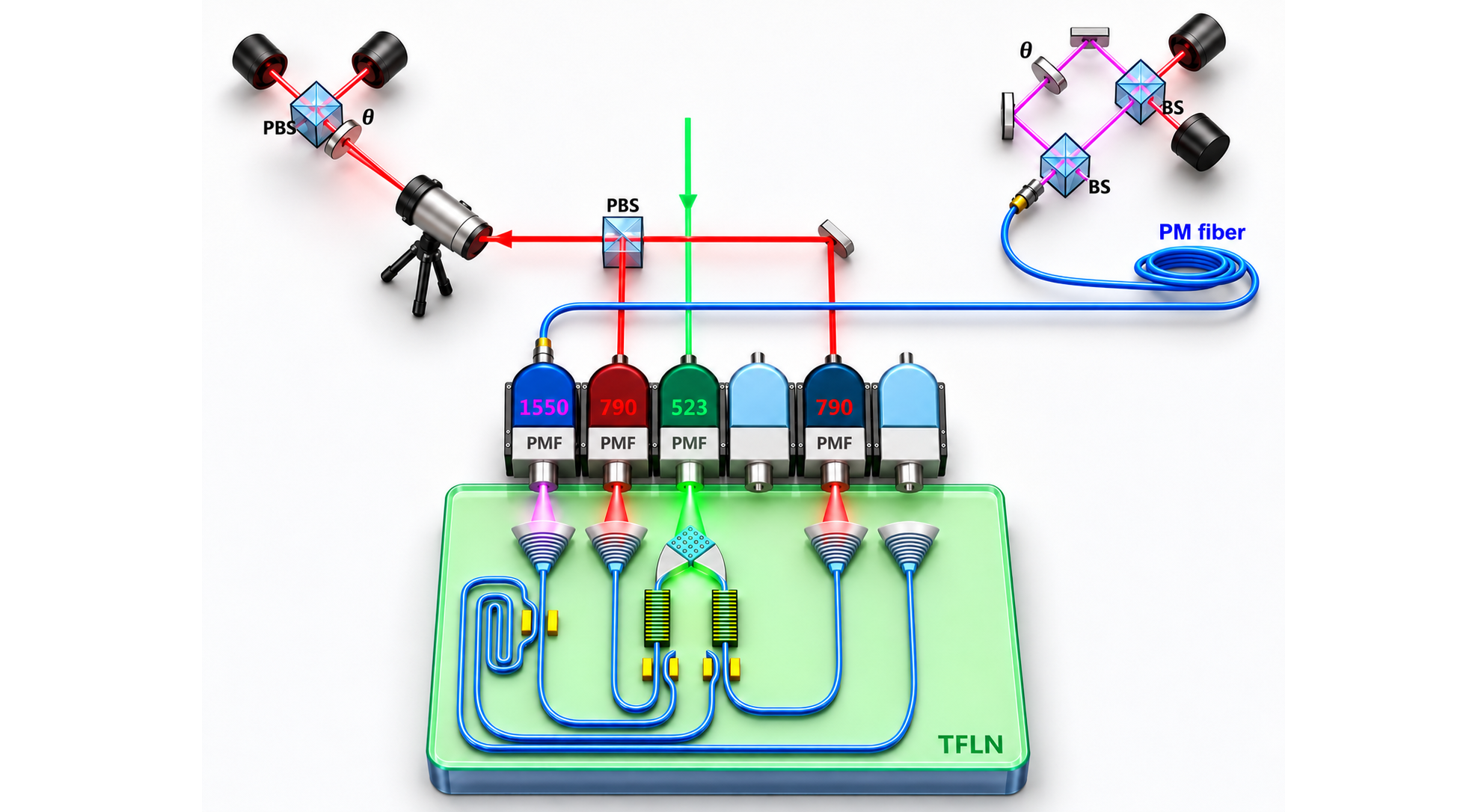}
	\footnotesize
	\caption{Schematic of the TFLN-based PIC for hybrid quantum entanglement. A PM fiber array is aligned to the chip to inject the $\SI{523.5}{\nano\meter}$ pulsed pump beam and collect two $\SI{790}{\nano\meter}$ signal outputs and one $\SI{1550}{\nano\meter}$ idler output.
}\label{HybridTFLNsetup}
\end{figure*}
The efficiency of EPSs on TFLN can be improved by several recent advances, including thicker films and deeper etches that tightly confine all interacting modes, edge couplers and photonic wire bonds for low-loss fiber interfacing, and cavity enhancements \cite{Lemieux2026,Franken2025}. Each of these approaches, however, introduces additional challenges in fabrication, packaging, or active stabilization. In this work, we deliberately take a different approach; namely, rather than maximizing efficiency, we impose a set of design constraints chosen for simplicity and robustness, and ask how far a TFLN-based PIC can support quantum communication under these constraints. The proposed architecture, shown in Fig.~\ref{HybridTFLNsetup}, monolithically integrates polarization-splitting and standard grating couplers for external optical interfacing, periodically poled waveguides for SPDC, wavelength-selective directional couplers (DC) for separating the signal and idler photons, and unbalanced optical paths for time-bin encoding. External optical interfacing is provided by a polarization-maintaining (PM) fiber array block comprising one input fiber for the $\SI{523.5}{\nano\meter}$ pump and four output fibers - two for the $\SI{790}{\nano\meter}$ signal outputs and two for the $\SI{1550}{\nano\meter}$ idler outputs.

\begin{figure}[htbp!]
	\centering
	\includegraphics[width=\columnwidth]{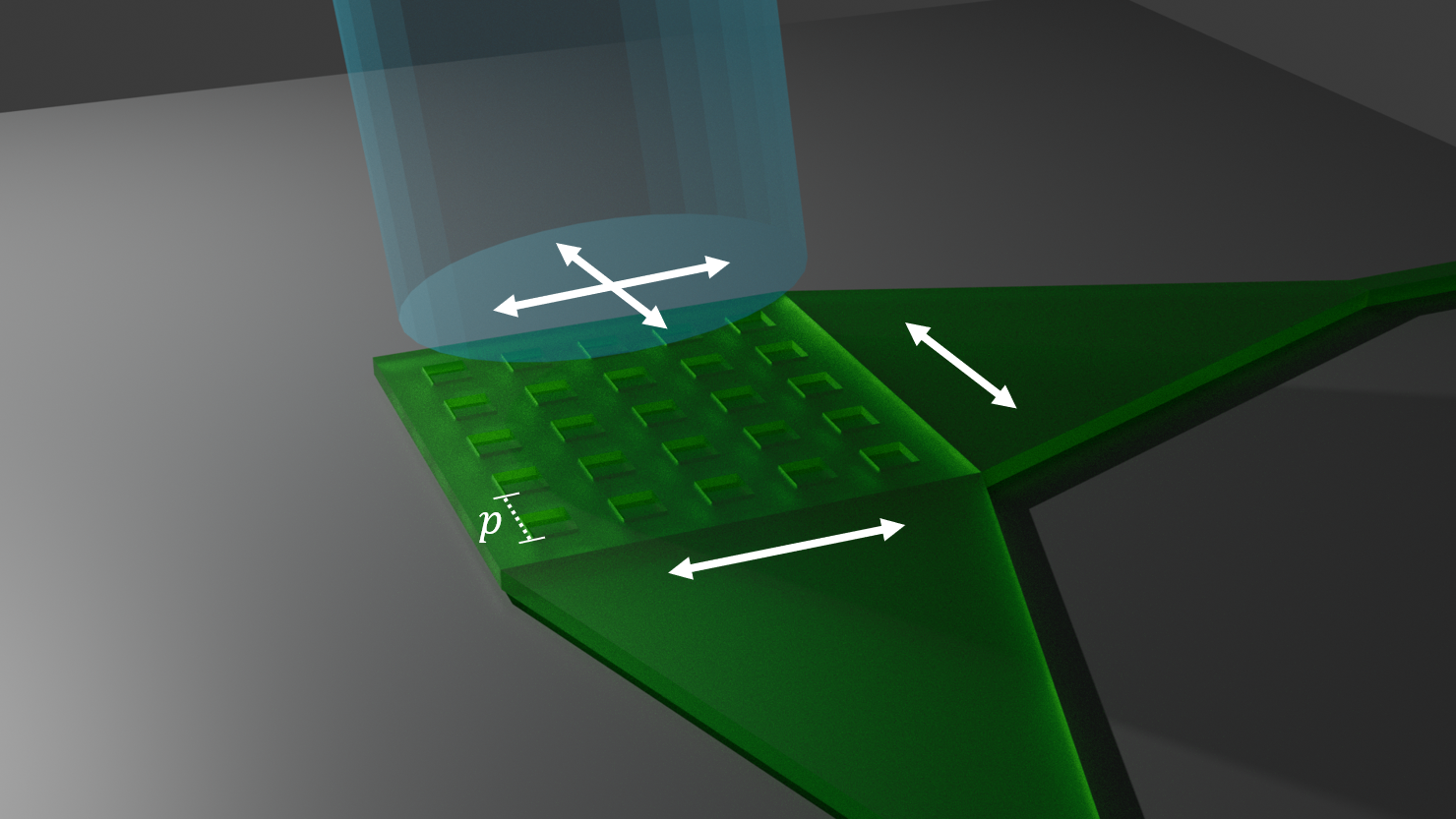}
	\footnotesize
	\caption{Polarization Splitting Grating Coupler (PSGC)
}\label{TFLN_PSGC_final}
\end{figure}

The pump is injected through a PM fiber that is coupled to the circuit through a polarization-splitting grating coupler (PSGC). As illustrated in Fig.~\ref{TFLN_PSGC_final} the PSGC splits the incoming pump beam into vertical and horizontal components, both of which enter the waveguide in the same TE mode. 
We adopt a rib waveguide geometry with a $\SI{150}{\nano\meter}$ film thickness and a $\SI{70}{\nano\meter}$ etch depth so that the $\SI{523.5}{\nano\meter}$ pump is single-mode in the vertical direction. This choice suppresses unwanted SPDC processes driven by vertical higher-order pump modes by construction, without requiring vertical mode-filtering elements, at the cost of reduced grating coupler efficiencies at $\SI{790}{\nano\meter}$ and $\SI{1550}{\nano\meter}$. While the film thickness and etch depth are fixed, the waveguide widths are varied locally along the circuit to optimize individual photonic components. Therefore, mode filtering in the lateral direction can still be done by tailoring the waveguide width. In each of the two waveguides branching from the PSGC, the filtered pump TE mode then propagates into its respective periodically poled $\SI{1}{\micro\meter}$-wide region, where a fraction of the pump power is converted into signal and idler photon pairs. Given the large spectral separation between the signal and idler photons, a directional coupler can separate the two photons in each path by exploiting the difference in their coupling lengths. From the FDTD simulations, the calculated coupling length is $\SI{29}{\micro\meter}$ for the $\SI{1550}{\nano\meter}$ idler photons and $\SI{274}{\micro\meter}$ for the $\SI{790}{\nano\meter}$ signal photons, indicating that the idler photons couple into the adjacent waveguide over a much shorter distance than the signal photons. This also offers a potential advantage over a Bragg grating, which typically suffers from higher optical loss \cite{Prencipe2021}. Microring resonators are also unsuitable here, as they restrict the idler photons to a narrow spectral bandwidth and can make the coherence length too long relative to the path-length difference required for time-bin entanglement. 
After spectrally separating the generated photon pairs, the $\SI{790}{\nano\meter}$ signal photons from both arms are coupled out into free space and recombined using a polarizing beam splitter (PBS) to achieve polarization-based entanglement. One of the signal arms can be rotated to the orthogonal polarization either by placing a half waveplate (HWP) before the PBS or simply by rotating the axis of its PM fiber output port. While coupling loss at the input PSGC is tolerable, as it only reduces the available pump power prior to SPDC, optical losses after photon-pair generation directly compromise pair transmission and should therefore be minimized. Consequently, on-chip recombination of the signal photons could be considered in the future once PSGC designs achieve significantly higher coupling efficiency.
In contrast, the $\SI{1550}{\nano\meter}$ idler photons exit through only one of the idler output ports in the Early (E) and Late (L) time bins. On the idler arms, an additional DC is used to recombine the two idler modes, forming an unbalanced interferometer together with the input PSGC. At the receiver, a matching unbalanced interferometer produces coherent EL and LE superposition, completing the time-bin entanglement.
We implement time-bin entanglement by making the two path lengths differ by $l=4cm$. This path-length difference corresponds to a time delay of $\Delta T\approx\SI{250}{\pico\second}$ according to the following equation
\begin{equation}
    \Delta T=\dfrac{n_{\text{eff,i}}l}{c}
    \label{eq:path-length}
\end{equation}
This delay is chosen to exceed the pump pulse width $\tau_p=\SI{50}{\pico\second}$ while remaining short enough to avoid overlap between consecutive pulses separated by $\SI{1}{\nano\second}$ ($f_{\text{rep}}=\SI{1}{\giga\hertz}$). It is worth noting that the time delay needs to be sufficiently large to also account for detector and electronic timing jitter. While hybrid entanglement has previously been demonstrated using bulk crystals and optics\cite{Fujiwara2009,Ma2009}, the TFLN-based PIC architecture presented here brings this capability onto a compact integrated platform for quantum networking.

%%%%%%%%%%%%%%%%%%%%%%%%%%%%

\section{Discussions and Outlook}
The EPS chip design presented here serves as a conservative baseline to assess the lower-bound performance achievable with a simplified PIC architecture for long-distance hybrid-channel quantum communication. As shown in Table \ref{tab:loss-budget}, the total loss is $\SI{9.00}{dB}$ for the signal and $\SI{18.67}{dB}$ for the idler. The higher loss in the idler path comes mainly from the $\SI{4}{\centi\meter}$ spiral delay line of the unbalanced interferometer. The pump insertion loss at the PSGC ($\SI{10.62}{dB}$) does not reduce the heralding efficiency directly, since it can be compensated by increasing the pump power.

\begin{table}[htbp!]
\setlength{\extrarowheight}{2pt}
\renewcommand{\arraystretch}{1.3}
\caption{\label{tab:loss-budget} Estimated optical loss budget for the hybrid entangled-photon source.}
\begin{ruledtabular}
\begin{tabular}{lccc}
\multirow{2}{*}{\textbf{Component}} & \multicolumn{3}{c}{\textbf{Loss (dB)}} \\
\cline{2-4}
 & \textbf{Pump} & \textbf{Signal} & \textbf{Idler} \\
\hline
PSGC (PMF $\to$ PIC) & $10.62$ & --- & --- \\
DC Filter\footnote{Idler photon crosses; signal photon remains.} & --- & $0.003$ & $0.81$ \\
Waveguide\footnote{Combined loss across all on-chip waveguide sections, including straight and bent regions.} & -- & $0.5$ & $5.16$ \\
GC (PIC $\to$ PMF) & --- & $8.50$ & $12.70$ \\
\hline
\textbf{Total loss} & \textbf{10.62} & \textbf{9.00} & \textbf{18.67} \\
\end{tabular}
\end{ruledtabular}
\end{table}
To put these numbers in context, the QEYSSat mission was designed to operate with a total link loss of about $\SI{43}{dB}$~\cite{Henri2026}. If the signal photon is sent through the free-space channel, the on-chip loss of $\SI{9.00}{dB}$ leaves only about $\SI{34}{dB}$ for the channel itself. Horizontal links over about $\SI{20}{\kilo\meter}$, such as mountain-to-mountain or building-to-building links, have shown channel losses of roughly $\SI{20}{dB}$~\cite{LopezGrande2026, Cai2024}, and drone-based entanglement distribution has been demonstrated with channel losses near $\SI{12}{dB}$~\cite{Liu2020}. Our EPS therefore fits within the loss budget of these platforms.
The main limitation of the design comes from the choice of a thin film ($\SI{150}{\nano\meter}$) and a shallow etch depth ($\SI{70}{\nano\meter}$).  This geometry was chosen so that the waveguide supports only the fundamental pump mode in the vertical direction; otherwise, the pump light coupled in through the grating coupler could also excite higher-order modes. However, this geometry also weakens the confinement of the signal and idler modes, especially at $\SI{1550}{\nano\meter}$, which leads to larger losses, mainly at the output grating couplers. This limitation could be removed with a mode-matched coupling method such as photonic wire bonding. In that case, the pump could be launched directly into its fundamental mode, so strict vertical mode filtering would no longer be needed. A thicker film could then be used and may greatly reduce the losses in the signal and idler modes.
Also, in this case, the pump could be split into two fundamental TE modes with a simple on-chip 50:50 splitter, such as a directional coupler, instead of the PSGC. Nonetheless, our simulation results show that some long-distance links are already possible even with the current design, which prioritizes simplicity. This is enabled by the strong second-order nonlinearity of TFLN for efficient photon-pair generation, making it a very promising platform for entanglement-based quantum communication. Even though PICs still suffer from higher optical loss than bulk free-space sources, optimized grating couplers and edge couplers on TFLN have reached coupling losses of a few dB or less~\cite{Chen2022,He2019}. Propagation losses well below $\SI{0.5}{dB/cm}$ have been reported in TFLN waveguides~\cite{Zhang2017}, and their performance is expected to improve further as fabrication and material technologies advance. With these potential improvements, the photon-pair loss could be greatly reduced, which would bring a satellite link within reach.

%%%%%%%%%%%%%%%%%%%%%%%%%%%%

\section{Acknowledgement}
This work is supported by the Natural Sciences and Engineering Research Council (NSERC) of Canada (ALLRP587216-23, ALLRP578460-22, and RGPIN-2024-05292), NSERC Alliance (R832825) and in part by the Canada Excellence Research Chair (CERC) Program (R832756).

%%%%%%%%%%%%%%%%%%%%%%%%%%%%

\onecolumngrid
\section{Appendix : FDE \& FDTD simulations}
\begin{figure}[ht]
\centering
	\centering
    \includegraphics[width=0.8\textwidth]{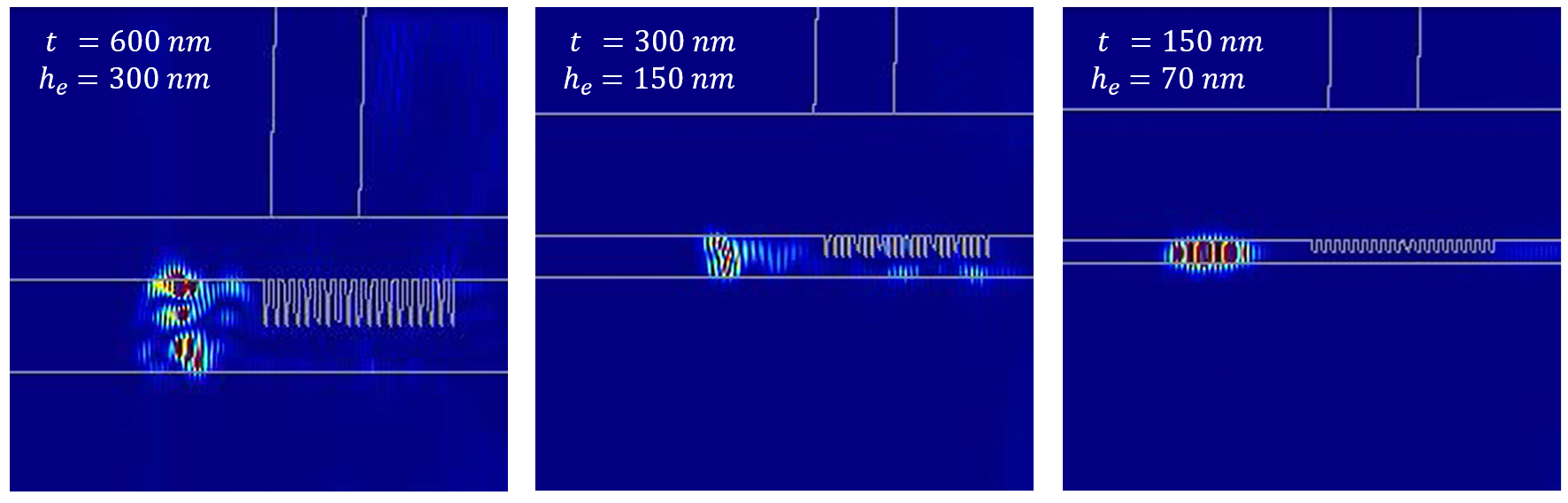}
	\footnotesize
	\caption{2D FDTD simulations of grating-coupler injection for the $\SI{523.5}{\nano\meter}$ pump across different TFLN film thicknesses $t$ and etch depths $h_e$.}\label{GC_523}\label{GC_523}
\end{figure}
\twocolumngrid

The parameter choice and the optical performance of each circuit component are evaluated here using the FDE and FDTD simulations.
As noted earlier, these choices favor simplicity over minimal optical loss, so the expected performance here should be read as a conservative baseline. Further improvements in component design or fabrication would reduce optical loss, potentially enabling longer-distance or harsher-environment quantum communication.

Supporting only a single vertical mode for the $\SI{523.5}{\nano\meter}$ pump requires a relatively thin TFLN layer. The V-number is given by 
\begin{equation}
\setlength{\jot}{10pt}  
	\begin{aligned}\label{kappa}
	V=\dfrac{\pi t}{\lambda}\sqrt{n_{wg}^2-n_{clad}^2},
	\end{aligned}
\end{equation}
where $t$ is the thickness of the waveguide, and $n_{wg}=2.237$ and $n_{clad}=1.461$ are the refractive indices at the pump wavelength. 
Cutting off the first odd mode in the vertical direction ($\mathrm{TE}_{01}$) requires $V<\pi/2$, which yields a cutoff thickness of $t_{cut}\approx\SI{155}{\nano\meter}$, as confirmed by our simulations in Fig.\ref{GC_523}. While the thickness and the etch depth are fixed across the entire circuit, the waveguide width can be chosen independently for each component. Determining the single-mode cutoff directly in a partially etched rib waveguide can give misleading effective indices, because slab-confined modes may accidentally appear as guided core modes. 
\begin{figure}[htbp!]
	\centering
	\includegraphics[width=\columnwidth]{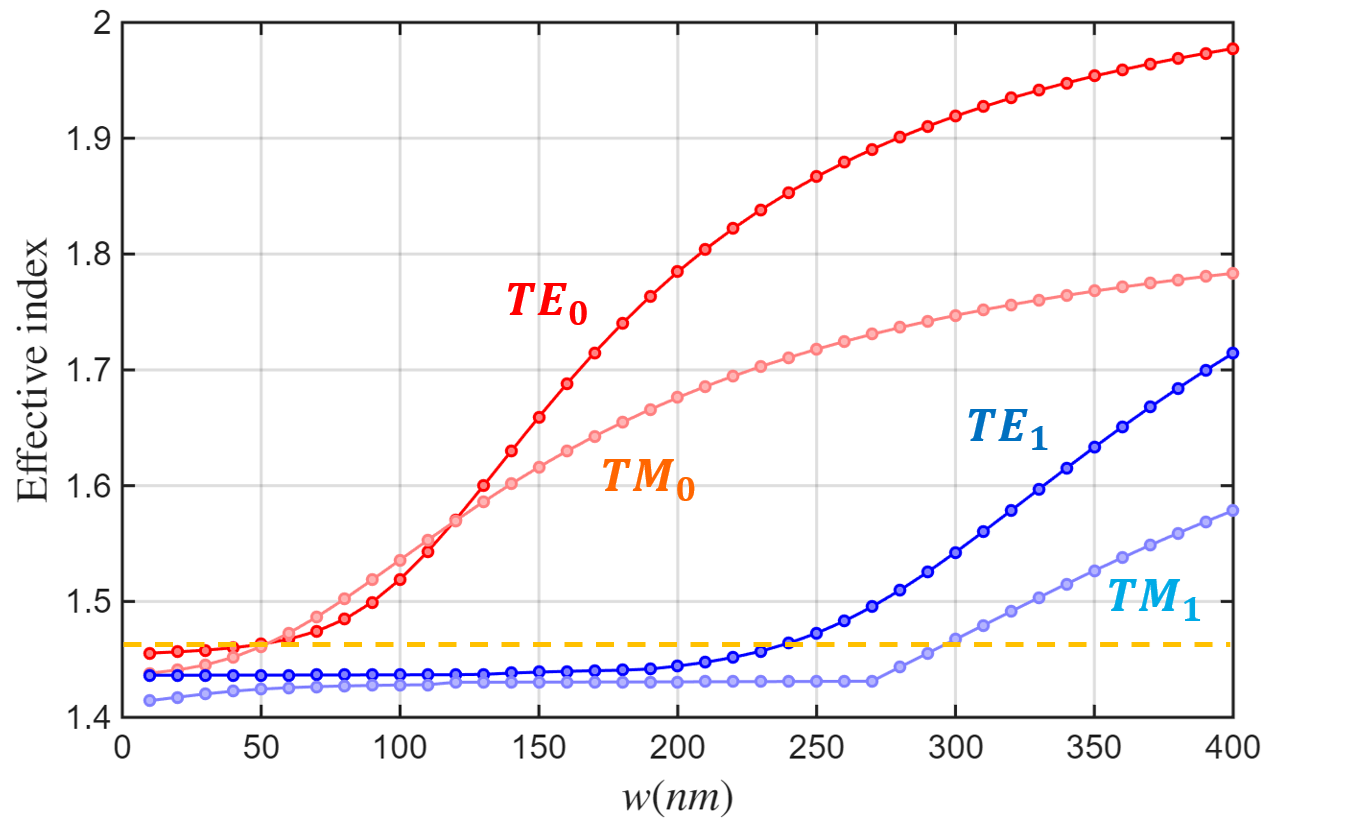}
	\footnotesize
	\caption{Effective refractive indices of the guided modes at $\SI{523.5}{\nano\meter}$ as a function of waveguide width $w$ for a fully etched waveguide ($t=\SI{150}{\nano\meter}$, $\ang{90}$ sidewall).}\label{N_eff_523_90deg}
\end{figure}

\begin{figure}[htbp!]
	\centering
	\includegraphics[width=\columnwidth]{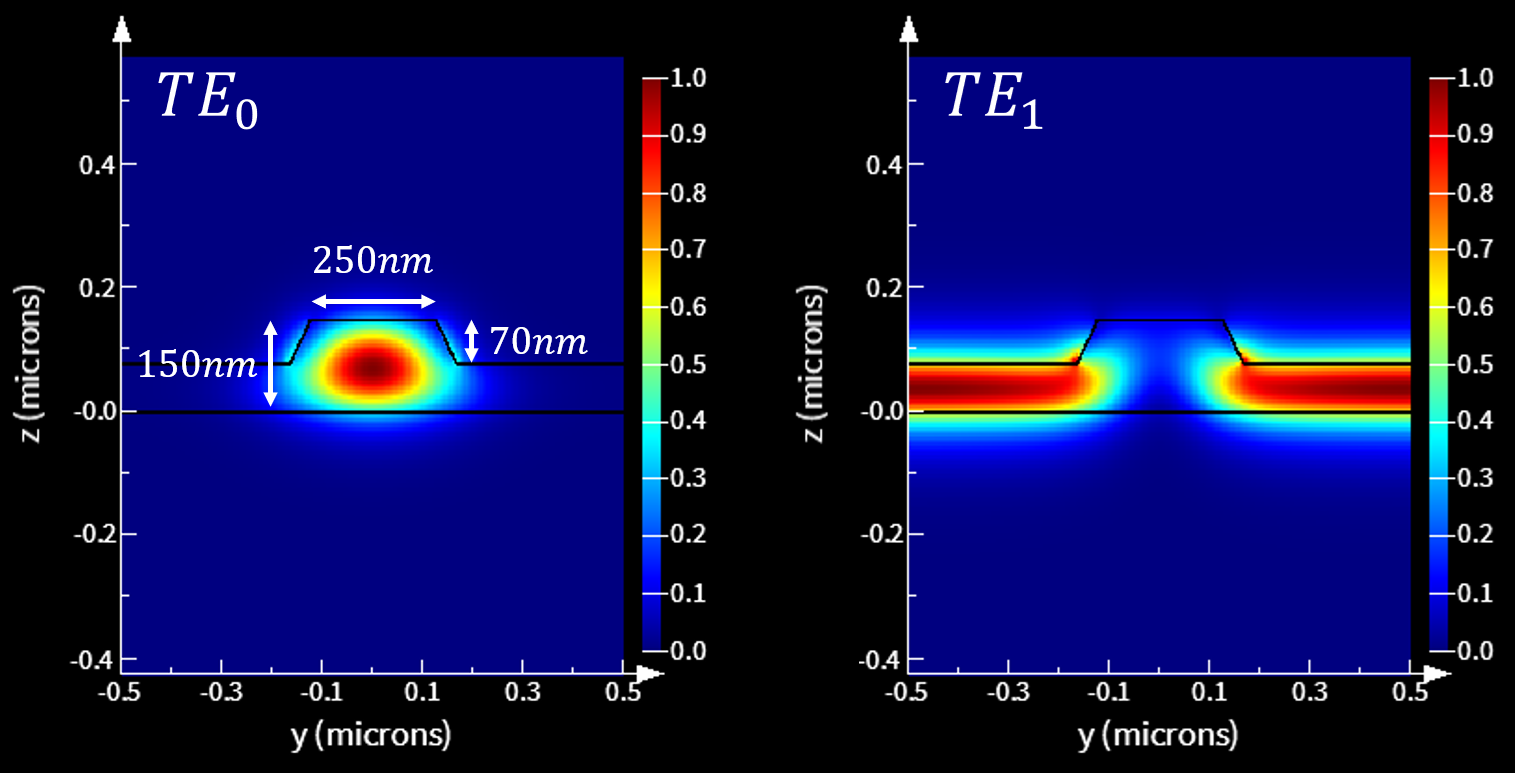}
	\footnotesize
	\caption{$\mathrm{TE}_0$ (left) and $\mathrm{TE}_1$ (right) pump mode profiles.}\label{523_ribWG}
\end{figure}

To avoid this, we first determine the single-mode width using fully etched waveguides, and then add the slab to obtain the effective indices of the rib waveguide used in our SPDC calculations. For instance,  to also cut off the first odd mode of the pump in the horizontal direction ($\mathrm{TE}_{10}$), when a rectangular (fully etched) waveguide is considered, the width of the waveguide tapered from the PSGC should be narrower than $\SI{250}{\nano\meter}$, as shown in Fig.~\ref{N_eff_523_90deg}. Achieving a vertical sidewall in a rib waveguide is challenging, and the actual sidewall angle can vary depending on the foundry's etching and fabrication capabilities. Currently, due to its chemical properties, TFLN is mostly patterned by physical etching, resulting in sloped sidewalls ($\ang{60}-\ang{75}$) \cite{Peng2025,Xing25,Chen26}. 
\begin{figure}[htbp!]
	\centering
	\includegraphics[width=\columnwidth]{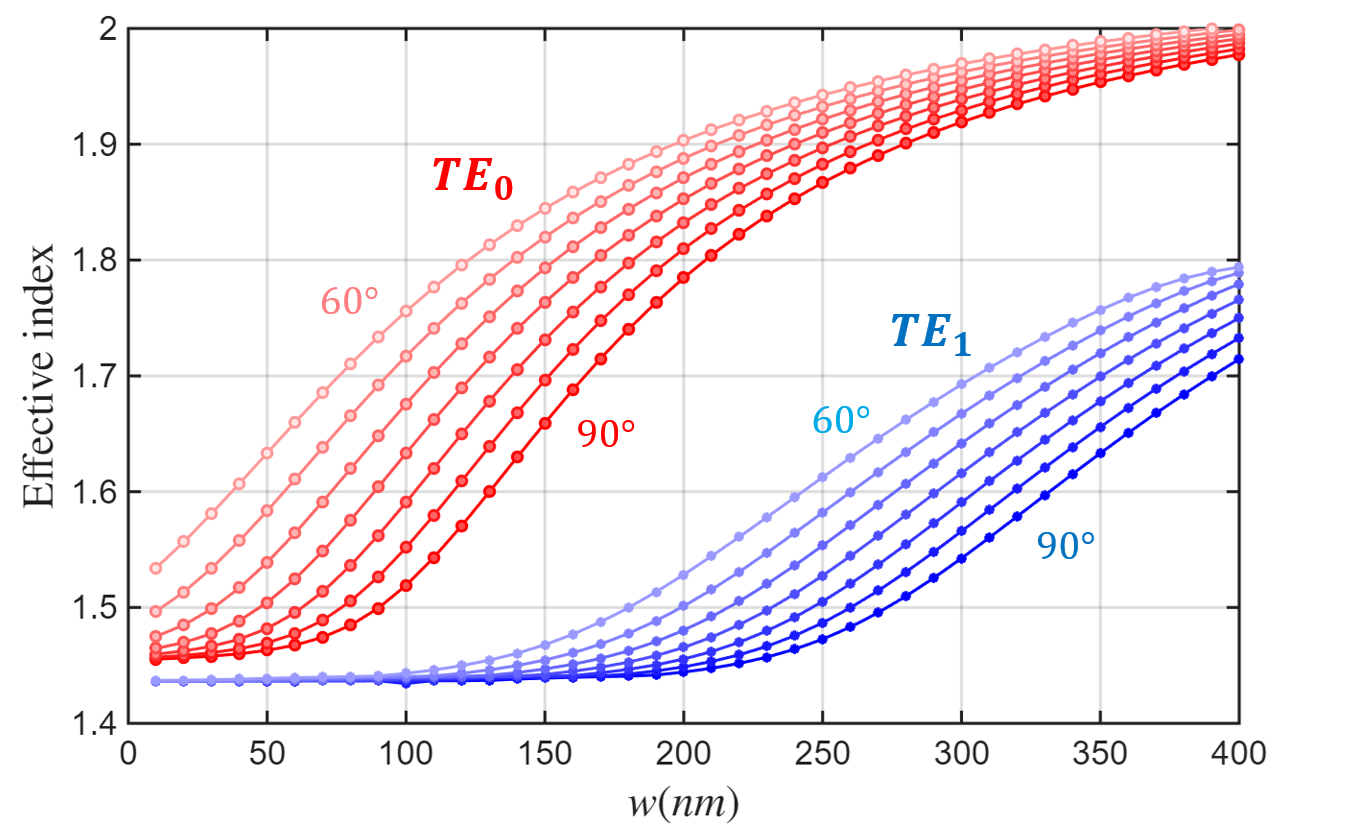}
	\footnotesize
	\caption{Following Fig.~\ref{N_eff_523_90deg}, effective indices of $\mathrm{TE}_0$ and $\mathrm{TE}_1$ with the sidewall angle varied from $\ang{60}$ to $\ang{90}$.}\label{N_eff_523_angles}
\end{figure}
As shown in Fig.~\ref{N_eff_523_angles}, a smaller sidewall angle increases the effective width of the waveguide, so a fully-etched waveguide would require a narrower top width to remain single-mode. In our rib waveguide, however, the remaining slab weakens the lateral confinement, which reduces the number of guided higher-order modes. According to our simulation in Fig.~\ref{523_ribWG}, a top width of $\SI{250}{\nano\meter}$ remains a reasonable choice to support only the TE$_0$ mode at the pump wavelength.

The coupling efficiency of the PSGC was calculated assuming input from a PM fiber designed for pump, with a core diameter of $\SI{3}{\micro\meter}$. As shown in Fig.~\ref{TFLN_PSGC_final}, the fast axis of the PM fiber is oriented diagonally ($\phi$=$\ang{45}$) with respect to the two grating axes of the PSGC. The input field is therefore projected equally onto the two orthogonal polarization components, each of which is diffracted by the two-dimensional grating into one of the two orthogonal tapered waveguides. To first order, the grating pitch $p$ along each waveguide axis is given by the phase-matching condition
\begin{equation}\label{pitch}
	p = \frac{\lambda}{n_\mathrm{eff} - n_\mathrm{c}\sin\theta\cos\phi},
\end{equation}
where $n_\mathrm{c}$ is the refractive index of the cladding and $\theta$ is the fiber tilt angle from the surface normal. Eq.~\eqref{pitch} provides an initial estimate, and the pitch and duty cycle $d$, defined as the ratio of the unetched hole width to the pitch, are subsequently optimized through numerical simulations to maximize the coupling efficiency. With the fiber tilt angle set to $\theta$=$\ang{8}$ to suppress Bragg reflection, the optimized PSGC has a pitch of $\SI{297}{\nano\meter}$ and a duty cycle of 0.77. The resulting coupling efficiency is $8.6$\%, corresponding to $4.3$\% coupled into each of the two orthogonal waveguides. Using the same film thickness and etch depth, we also designed standard grating couplers to couple the signal and idler photons out of the chip. The simulated coupling efficiencies are $14$\% for the signal and $5.4$\% for the idler.

After each SPDC region, we use a DC to separate the signal ($\SI{790}{\nano\meter}$) and idler ($\SI{1550}{\nano\meter}$) photons, so the DC acts as a spectral filter. In the coupling region, the two waveguides are both $\SI{1}{\micro\meter}$ wide and are separated by a gap of $g$=$\SI{500}{\nano\meter}$. Using FDE simulations, we obtain the effective indices of the even and odd supermodes. The coupling strength $\kappa$ is then given by half the difference between their propagation constants:
\begin{equation}\label{kappa}
	\kappa = \frac{|\beta_\mathrm{even}-\beta_\mathrm{odd}|}{2}.
\end{equation}
A fraction of power transferred into the other waveguide after a length $L$ is $P=\sin^2(\kappa L)$, so the coupling length needed for full transfer is $L_c=\pi/(2\kappa)$. The $\SI{1550}{\nano\meter}$ idler photon has a longer wavelength, so its mode extends further outside the waveguide and couples much more strongly to the neighboring waveguide than the $\SI{790}{\nano\meter}$ signal photon. We use this difference to transfer the idler photons into the other waveguide while keeping the signal photons in the original one. This requires $L_{c,i} \ll L_{c,s}$, so that the coupler length can be set to $L_{c,i}$ with little coupling of the signal. 
\begin{figure}[htbp!]
	\centering
    \includegraphics[width=\columnwidth]{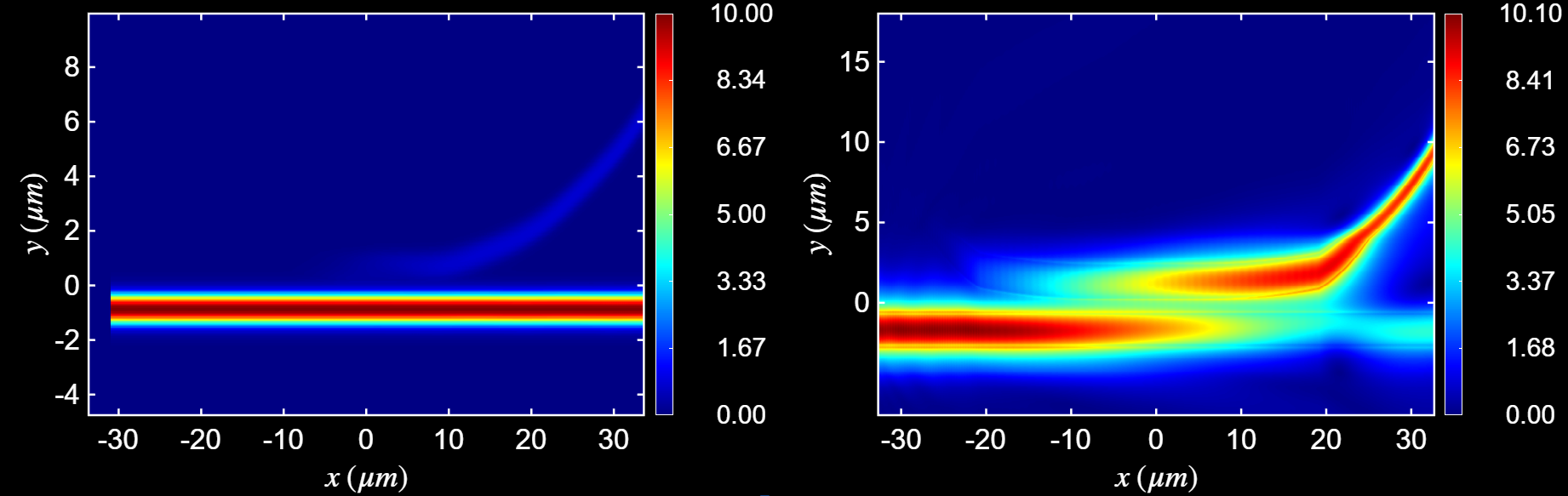}
	\footnotesize
	\caption{FDTD simulations of the directional coupler filter for $\SI{790}{\nano\meter}$ (left) and $\SI{1550}{\nano\meter}$ (right).}\label{DC_filter}
\end{figure}
From our FDE simulations, the coupling lengths are $L_{c,1550} = \SI{29}{\micro\meter}$ and $L_{c,790} = \SI{274}{\micro\meter}$, so the idler couples nearly an order of magnitude faster than the signal. However, this assumes coupling only in the straight section. In practice, the waveguides also couple in the bent regions where they approach and separate from each other, so the straight section required for full transfer is shorter. We therefore used $L_c$ as an upper bound and initial value, and determined the final length from full-device FDTD simulations. As shown in Fig.~\ref{DC_filter}, with a straight section of only $\SI{7}{\micro\meter}$ and a bend radius of $\SI{30}{\micro\meter}$, about $83$\% of the $\SI{1550}{\nano\meter}$ idler is transferred to the other waveguide, while $99.4$\% of the $\SI{790}{\nano\meter}$ signal remains in the original waveguide.

\twocolumngrid
%%%%%%%%%%%%%%%%%%%%%%%%%%%%

\section{References}
\bibliography{Paper/8-Reference}

\end{document}